\documentclass[twocolumn,showpacs,preprintnumbers,floatfix,psfig]{revtex4}
\usepackage{graphicx}
\usepackage{dcolumn}
\usepackage{bm}

\begin{document}

\title{Coulomb crystallization in two-component plasmas}

\author{M.~Bonitz$^{1}$, V.S.~Filinov$^{1,2}$, V.E.~Fortov$^{2}$, P.R.~Levashov$^{2}$, and H.~Fehske$^{3}$}

\affiliation{$^{1}$Institut f{\"u}r Theoretische Physik und Astrophysik, Christian-Albrechts-Universit{\"a}t zu Kiel, Leibnizstr. 15, 24098 Kiel, Germany}
\affiliation{$^{2}$Institute for High Energy Density, Russian Academy of Sciences,
Izhorskay 13/19, Moscow 127412, Russia}
\affiliation{$^{3}$Institut f\"ur Physik, Ernst-Moritz-Arndt-Universit{\"a}t
Greifswald, Domstrasse 10a, D-17489 Greifswald, Germany}
\date{\today}

\begin{abstract}
The analysis of Coulomb crystallization is extended from one-component to two-component  plasmas. Critical parameters for the existence of Coulomb crystals are derived 
for both classical and quantum crystals. In the latter case, a critical mass ratio of the two charged components is found which is of the order of 80. Thus, holes in semiconductors with sufficiently flat valence bands are predicted to spontaneously order into a regular lattice. Such hole crystals are intimately related to ion Coulomb crystals in white dwarf and neutron stars as well as to ion crystals produced in the laboratory. A unified phase diagram of two-component Coulomb crystals is presented and is verified by first-principle computer simulations.
\end{abstract}

\maketitle

Crystallization is one of the most fundamental many-particle phenomena in charged particle systems. After the prediction of a highly correlated state of the electron gas -- the electron Wigner crystal \cite{wigner} -- there has been an active search for this phenomenon in many fields. Crystallization of electrons was observed on the surface of helium droplets \cite{grimes} and is predicted to occur in semiconductor quantum dots \cite{afilinov-etal.01prl}. Moreover, crystals of ions have been observed in traps \cite{itano98} and storage rings \cite{habs01}, and are expected to occur in layered systems \cite{kalman1}. The necessary condition for the existence of a crystal in these one-component plasmas (OCP) is that the mean Coulomb interaction energy, 
$e^2/{\bar r}$ (${\bar r}$ denotes the mean inter-particle distance), exceeds the mean kinetic energy (thermal energy $\frac{3}{2}k_BT$ 
or Fermi energy $E_F$ in classical or quantum plasmas, respectively) by a factor 
$\Gamma$ larger than 
$\Gamma^{cr}$ which, in a classical OCP is given by $175$ \cite{grimes,dewitt}. In a quantum OCP at zero temperature the coupling strength is measured by the Brueckner parameter, 
$r_s\equiv {\bar r}/a_B$ ($a_B$ denotes the effective Bohr radius), with a 
critical value for crystallization of $r_s^{cr}\approx 100$ \cite{ceperley80}.

The vast majority of Coulomb matter in the Universe, however, exists in the form of {\em neutral plasmas}, containing (at least) two oppositely charged components (two-component plasma, TCP). Coulomb crystallization in a TCP  has been observed 
as well, e.g. in colloidal and dusty plasmas \cite{thomas94,arp04} and it is predicted to be 
possible in laser-cooled expanding plasmas \cite{pohl04}. The lattice of heavy particles is immersed into a structureless gas of the light component which does not affect the former. Besides these {\em classical TCP crystals} it is expected that 
in the interior of white dwarf stars and in the crust of neutron stars 
there exists an entirely different type of TCP crystals \cite{segretain}: Crystals of 
highly charged ions (e.g. fully ionized carbon, oxygen, iron) which are embedded into an extremely dense degenerate Fermi gas of electrons. No such {\em quantum TCP crystals} have been observed in the laboratory, despite early suggestions \cite{rice68}. 
It is an open question what classical and quantum TCP crystals (being 
separated by $15\dots 20$ orders of magnitude in density) have in common and if there  
exists an integrative phase diagram. 

This Letter aims to answer these questions.
We show that, in fact, a common phase diagram of Coulomb crystals in a  generic 
neutral TCP (consisting of electrons and point-like ions \cite{point}) exists which is governed by five parameters -- density and temperature (as in the OCP case) and, additionally, by the asymmetry of the heavy (h) and electron (e) components with respect to three  fundamental properties: Mass, charge and temperature, $M=m_h/m_e$, $Z=Z_h/Z_e$ and $\Theta=T_e/T_h$. We show that classical TCP crystals require a critical charge 
ratio $Z$ whereas quantum TCP crystals exist only if the mass ratio $M$ exceeds a critical value of about $80$.
As a consequence, we predict the existence of quantum TCP crystals of protons in a dense hydrogen plasma and crystals of holes in semiconductors. 

Let us consider a locally neutral system of electrons and $Z$-fold 
charged heavy particles. The stationary 
state of the TCP is characterized by the dimensionless electron temperature 
$T_e = 3 k_BT/2 E_B$ and mean inter-electron distance $r_{se}={\bar r}_e/a_B$, 
where $E_B$ denotes the e-h binding energy, and the 
dimensionless density is given by $na_B^3=3/(4\pi r^3_{se})$. 

{\it Classical crystal.} For the existence of a Coulomb crystal in the presence of 
a classical gas of electrons, we first require that the heavy component is able to form a 
classical  OCP crystal, i.e.
\begin{eqnarray}
\Gamma_h  & \ge & \Gamma^{cr},
\label{cc}
\end{eqnarray}
and, secondly, that the electrons do not destroy that crystal, e.g., as a result of  screening of the heavy particle interaction. However, the main obstacle for the crystal turns out to be the formation of e-h bound states (atoms, excitons etc.), because this drastically reduces the h-h correlation energy causing violation of  condition (\ref{cc}). Therefore, we require that no significant fraction of heavy particles is trapped in Coulomb bound states, for which a conservative estimate is given by $T_e\ge 1 $, stating that the electrons have sufficiently high kinetic energy to escape the binding potential $V_{eh}$. Making use of charge neutrality, $n_e=n_h Z$, and the definitions of 
$\Gamma$ and $T_e$ we find from Eq.~(\ref{cc}) that the classical TCP Coulomb crystal exists between the temperatures
$T^{(1)}_e$ and $T^{(2)}_e$, given by
\begin{eqnarray}
T^{(1)}_e \approx 1 \quad \mbox{and} \quad 
T^{(2)}_e(r_{se}) = \frac{3\Theta Z^{2/3}}{\Gamma^{cr}}\frac{1}{r_{se}},
\label{t1t2}
\end{eqnarray}
and only if the ion charge exceeds a critical value,
$Z^{cr}_e(\Theta,r_{se}) = \left(\Gamma^{cr}r_{se}/3\Theta\right)^{3/2}$,
which is independent of the mass ratio $M$. 

{\it Quantum TCP crystal and critical mass ratio.} In the presence of quantum electrons, 
the condition for crystallization of the heavy particles follows from the quantum OCP result, $r_{sh} \ge r_s^{cr}$. To be specific, we will concentrate on hydrogen-like Coulomb bound states, where
$E_B  =  \frac{Ze^2}{4\pi\epsilon_0\epsilon_r}\frac{1}{2a_B}$, and
$a_B =  \frac{\hbar^2}{m_r}\frac{4\pi\epsilon_0\epsilon_r}{Ze^2}$
[$\epsilon_r$ and $m_r$ are the background dielectric constant 
and the reduced mass $m_r^{-1}=m_h^{-1}(1+M)$].
Since $r_s^{cr}$ refers to the critical interparticle distance in units of the hydrogenic $a^H_B$ \cite{ceperley80}, we first transform to the relevant 
effective Bohr radius $a_B$. Further, we eliminate $r_{sh}$ by expressing ${\bar r}_h$ by ${\bar r}_e$ using again charge neutrality. Then the OCP result can be rewritten as
\begin{equation}
Z^{4/3}(M+1)r_{se}\ge r_s^{cr}.
\label{zq}
\end{equation}
This crystal of heavy particles will survive in the presence of electrons only if,  as in the classical case, bound states are unstable which, at zero temperature, 
occurs due to pressure ionization at densities above the Mott 
density, i.e. ${r_{se}} \le r_s^{{\rm Mott}} \approx 1.2$ (see below). With increasing temperature, ionization becomes possible at lower density which we indicate by a 
monotonically decreasing function $1/r_s^{{\rm Mott}}(T_e)$ which 
vanishes when $T_e \rightarrow 1$ because there thermal ionization prevails.
Combination of (\ref{zq}) with the existence of pressure ionization allows us to 
eliminate $r_{se}$, yielding the criterion for the existence of a TCP crystal in the presence of a neutralizing background of quantum electrons as 
\begin{eqnarray}
M \ge M^{cr}(Z, T_e) = \frac{r_s^{cr}}{Z^{4/3}\,r_s^{{\rm Mott}}(T_e)} - 1,
\label{mcr}
\end{eqnarray}
which exists in a finite electron density range between
\begin{eqnarray}
n^{(1)}(T_e) &=&\frac{3}{4\pi}\left[\frac{1}{r_{s}^{\rm Mott}(T_e)}\right]^3,
\;
n^{(2)}(T_e) = n^{(1)}(T_e)K^3,
\nonumber
%\label{nmin}
\\[1ex]
K &=& (M+1)/(M^{cr}+1),
\label{nmax}
\end{eqnarray}
and below a critical temperature $T^*$. The density limits follow from the Mott 
criterion and from Eq. (\ref{zq}), whereas 
$T^*$ is estimated \cite{afilinov-etal.01prl} 
by the crossing point of the classical and quantum asymptotics of an OCP 
crystal (\ref{cc}) and (\ref{zq})
\begin{eqnarray}
T^{*} = 6 \frac{Z^{2}\Theta (M+1)}{\Gamma^{cr}\,r_s^{cr}}.
\label{tstar}
\end{eqnarray}
In absolute units, the density interval scales as $Z^3$ and the critical temperature 
as $M \cdot Z^{4}\cdot \Theta$. The critical mass ratio for singly charged ions in an isothermal TCP ($\Theta=1$) equals 83 and decreases with increasing temperature 
(due to the lower Mott density) and with increasing $Z$.

{\em Examples}  ($\Theta=1)$. 
(i) For a crystal of ${\rm C}^{6+}$ [${\rm O}^{8+}$]
ions expected to exist in the interior of white dwarf stars, 
the minimum density is given by
$n^{(1)}_e(0)=2 \cdot 10^{26}{\rm cm}^{-3}$ [$6.6 \cdot 10^{26}{\rm cm}^{-3}$],
and $T^{*}=  10^{9}K$  [$4.2 \cdot 10^{9}K$] \cite{nmax_astro}. 
(ii) Hydrogen and helium are predicted to form crystals as well: A crystal of protons 
[$\alpha-$particles] is stable between
$n^{(1)}_e(0)= 0.9  \cdot 10^{24} {\rm cm}^{-3}$ $[7.5 \cdot 10^{24}{\rm cm}^{-3}]$
and $n^{(2)}_e(0)= 10^{28}{\rm cm}^{-3}$ $[5 \cdot 10^{30}{\rm cm}^{-3}]$,
and below $T^{*}= 6.6\cdot 10^4K$  [$4.2 \cdot 10^6K$] and should be achievable in laboratory 
experiments with laser or ion beam compression techniques.
(iii) Crystallization of holes in semiconductors ($Z=1$)
is predicted for materials with a hole to electron mass ratio 
$M\ge M^{cr}\approx 83$ \cite{abrikosov}.
This value is feasible in intermediate valence semiconductors, such as Tm[Se,Te]
\cite{wachter91}. For example, for $M=100$ (using $\epsilon_r=20$) the parameters are $n^{(1)}_e(0)=1.2\cdot 10^{20}{\rm cm}^{-3}$,
$n^{(2)}_e(0)=2.1 \cdot 10^{20}{\rm cm}^{-3}$  and $T^{*} \approx 9K$.

\begin{figure}[h]
\includegraphics[width=3.5cm, height=2.5cm,clip=true]{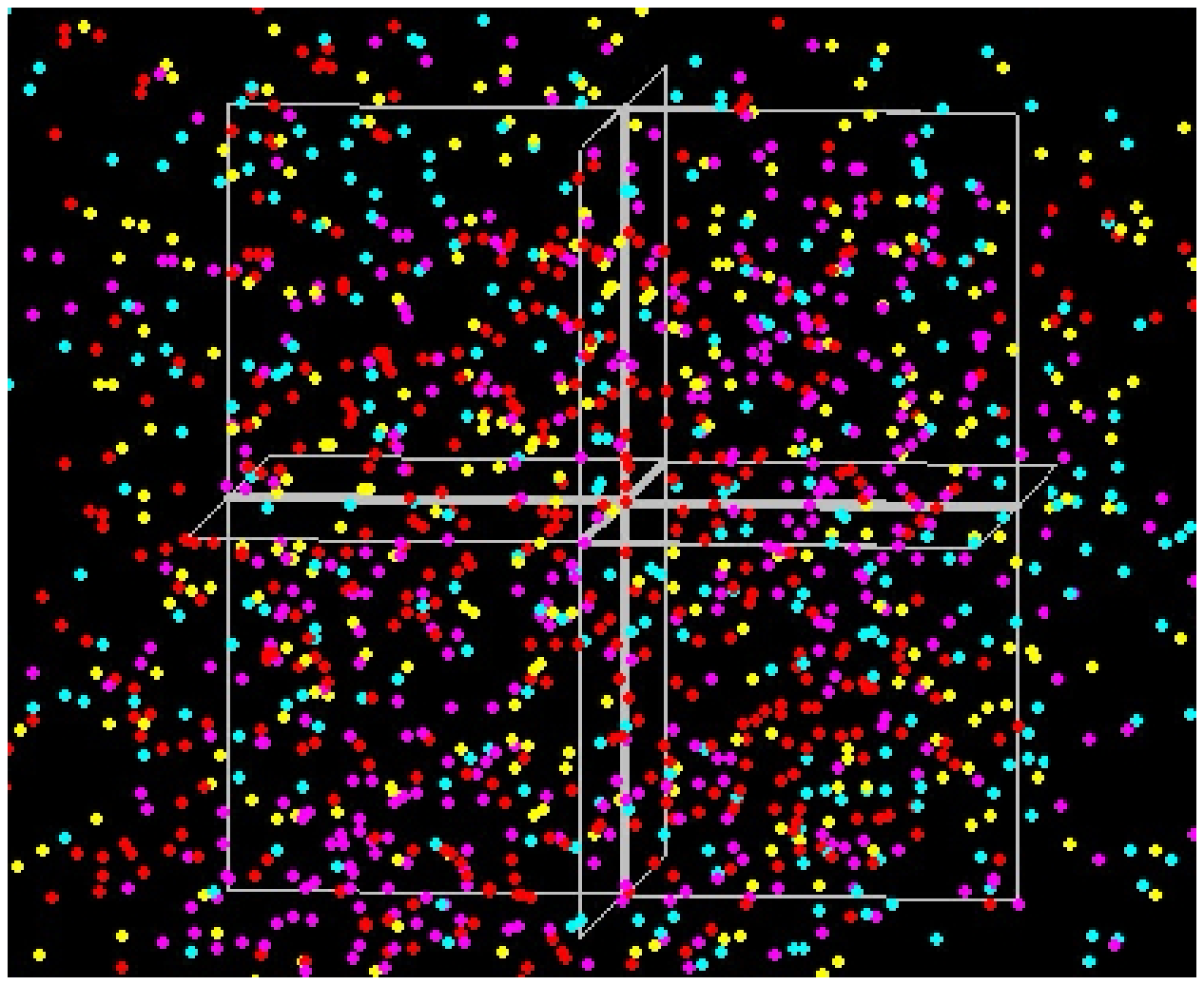}
\hspace{0.1cm}
\includegraphics[width=3.5cm,height=2.5cm,clip=true]{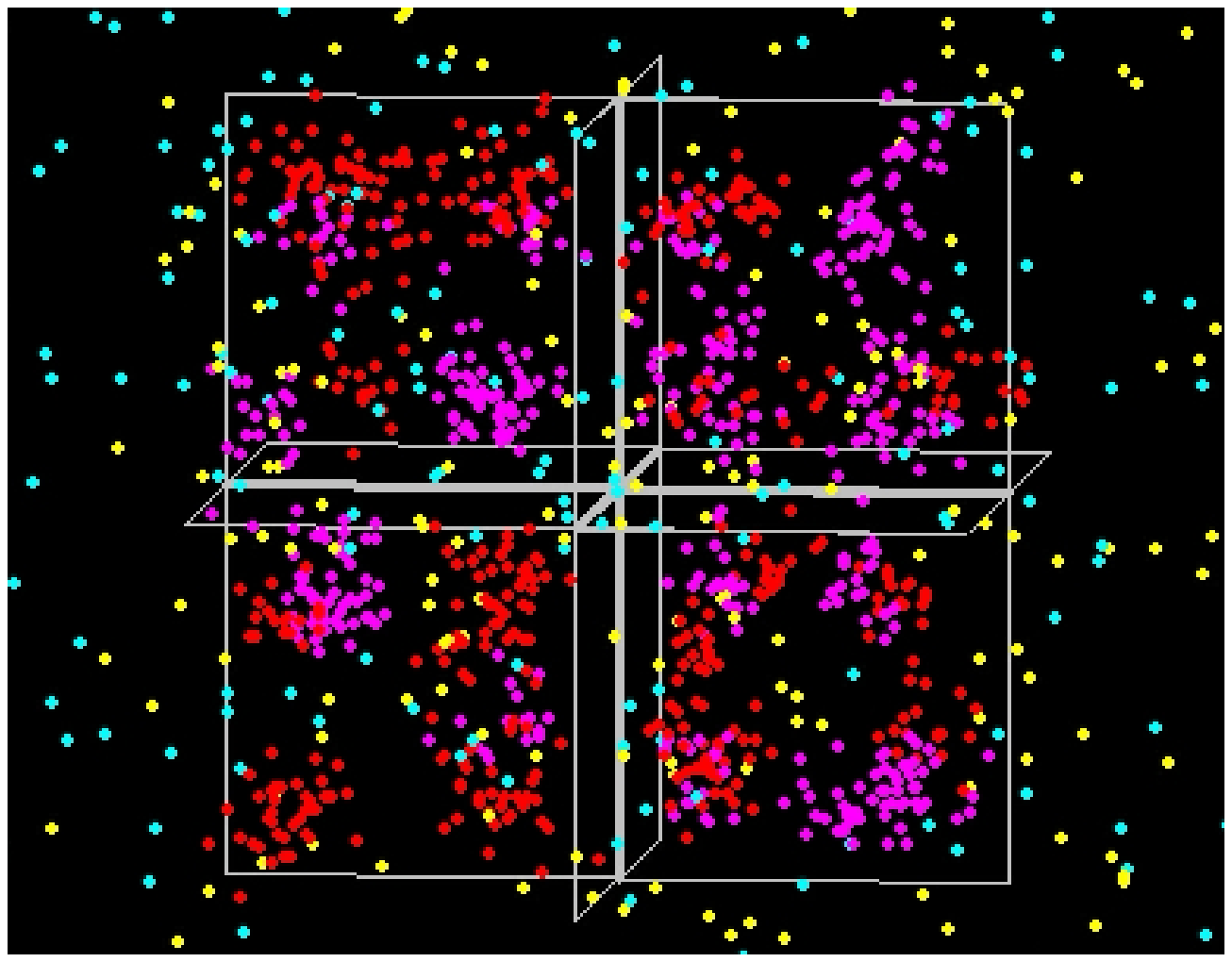}

\vspace{0.15cm}
\includegraphics[width=3.5cm,height=2.5cm,clip=true]{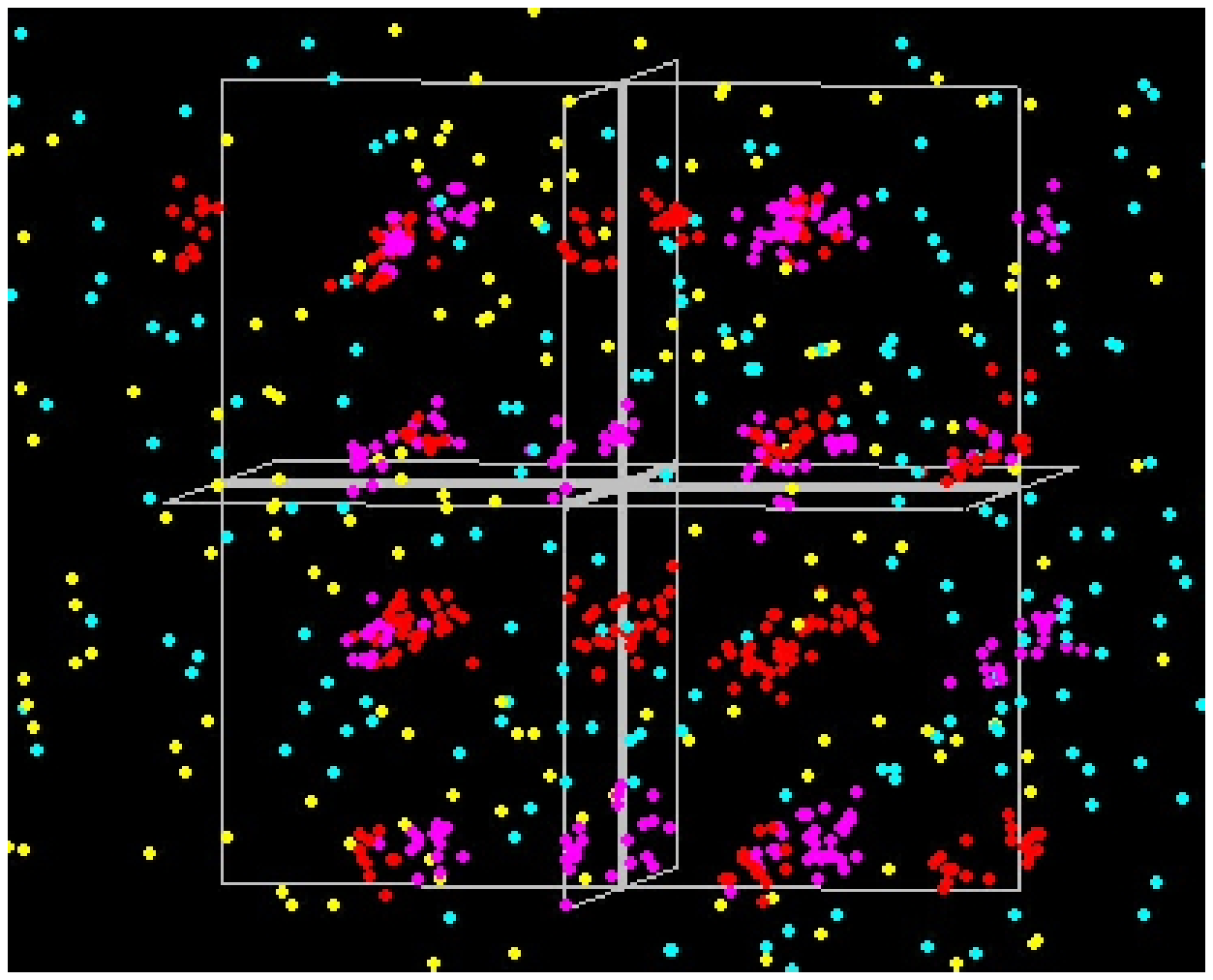}
\hspace{0.1cm}
\includegraphics[width=3.5cm,height=2.5cm,clip=true]{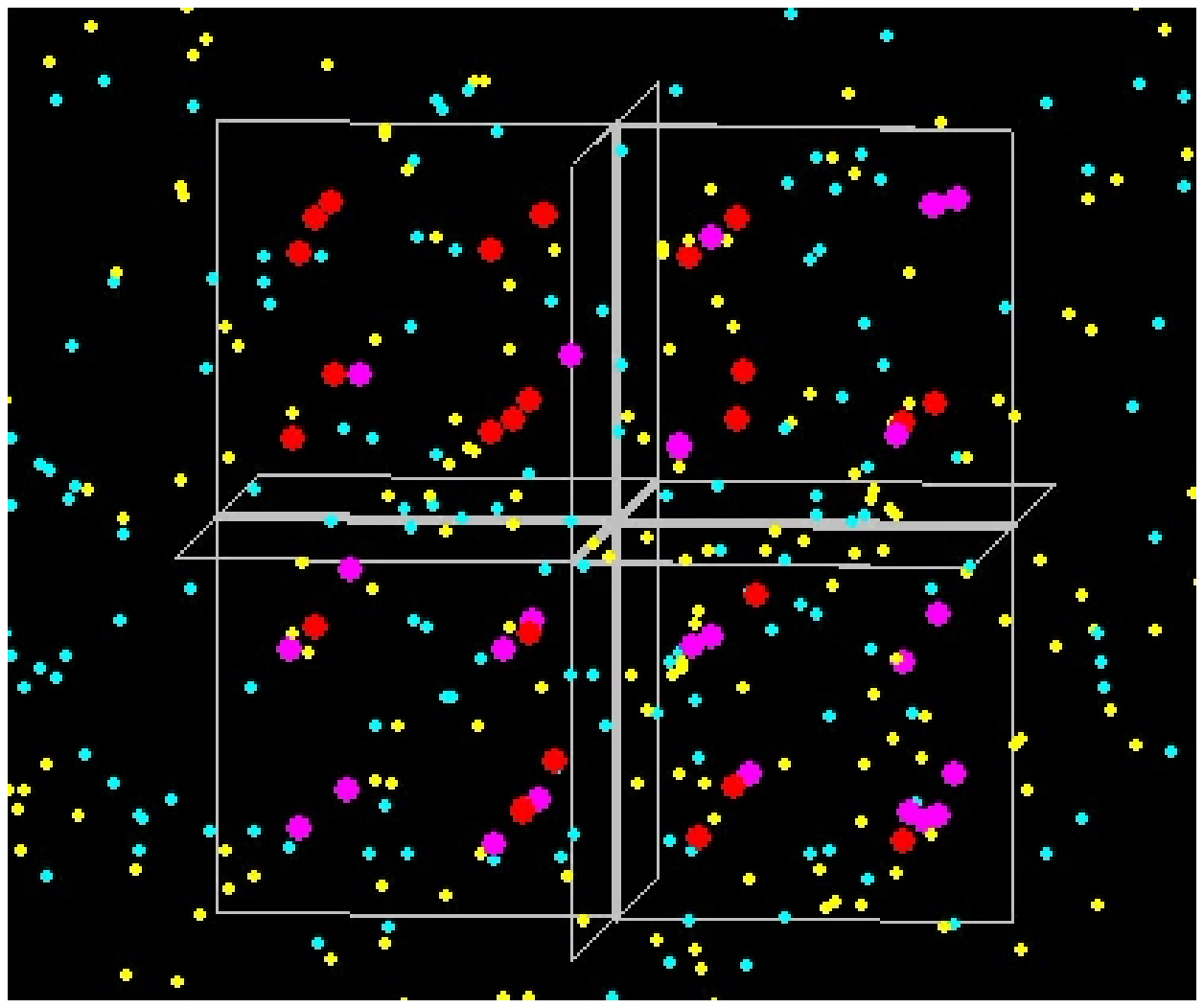}
\vspace{-0.3cm}
\caption{Snapshots of an e-h plasma at $T_e = 0.096$, and $r_{se}=0.63$.
Clouds of blue (yellow) dots mark the fully delocalized electrons with spin up (down), clouds of red (pink) dots denote holes with spin up (down).
Top left (right): $M=5$ $(50)$,
bottom left (right): $M=100$ $(800)$.}
\label{snap}
\end{figure}

{\it Numerical verification.} Of course, the boundary of the TCP crystal phase at 
low densities contains some uncertainty, due to the very complex transition from an  atomic system to a Coulomb crystal of heavy particles embedded into delocalized electrons. This transition which extends over a finite density interval may involve liquid-like behavior,  clusters and, at low temperature, phase separation, an analysis of which is beyond the present study. We estimate that these effects give rise to an uncertainty of the minimum density (Mott density), $n^{(1)}$, of the order of $30\%$. Further, the error of $r_s^{cr}$ is about $20\%$ \cite{ceperley80}, thus 
the critical parameters carry an uncertainty of about $50\%$. For particular systems, more accurate predictions are possible if the Mott parameter $r_s^{{\rm Mott}}$ is known, e.g. from computer simulations. Note that the complex processes of interest pose an extreme challenge to the simulations: They must self-consistently include the full Coulomb interactions, e-h bound state formation in the presence of a surrounding plasma, pressure ionization and the quantum and spin properties of the light and heavy species. 

\begin{figure}[h]
\includegraphics[width=7.cm,height=6cm, clip=true,angle=-0]{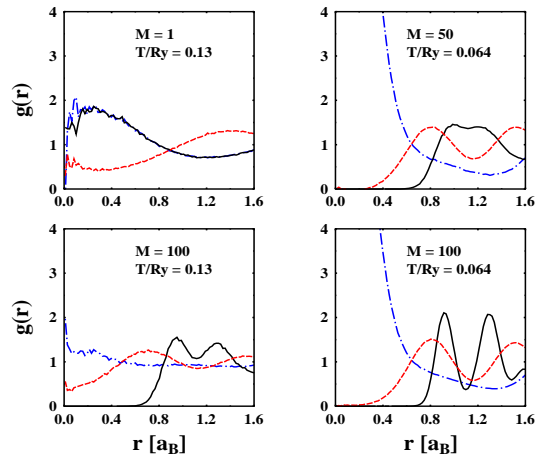}
\vspace{-0.35cm}
\caption{(Color online) e-e (blue, dash-dots), h-h (black, full line) 
and e-h (red, dotted line) pair distribution functions for different
mass ratios and temperatures.}
\label{pdf} 
\end{figure}

We therefore have performed extensive direct fermionic path integral 
Monte Carlo (PIMC) simulations of electron-hole plasmas which are based on our previous results for dense hydrogen-helium plasmas \cite{filinov-etal.01ppcf}, e-h plasmas \cite{filinov-etal.03jpa} and electron Wigner crystallization \cite{afilinov-etal.01prl}. While the so-called sign problem prohibits PIMC simulations of the ground state of a fermion system, here we 
restrict ourselves to temperatures at the upper boundary of the hole crystal phase, 
i.e. $T_e=0.06\dots 0.2$.
Studying mass ratios in the range of $M=1\dots 2000$ and densities corresponding to $r_{se}=0.6 \dots 13$ the simulations cover a large variety of Coulomb systems -- from positronium, over typical semiconductors to hydrogen. 
We start with the case of low densities (large $r_{se}$) to determine the Mott density $n^{(1)}$. Here the 
TCP consists of excitons and biexcitons, and we   
found \cite{fehske-etal.05jp} that for $r_{se}\le 1.2$ less than $10\%$ of the electrons and holes are bound, approving 
the choice of $r_s^{{\rm Mott}}$ made above.	
Thus crystallization should become possible. This is confirmed by our simulations,  see Figs.~\ref{snap} and \ref{pdf}, showing results for $r_{se}=0.63$ and different mass ratios $M$. Fig.~\ref{snap} displays snapshots of the e-h system in the simulation box. In all figure parts, the electrons form a nearly homogeneous Fermi gas -- individual electrons penetrate each other, extending far beyond the main simulation cell (shown by the grey grid lines). At the same time, the hole arrangement changes dramatically. While, at $M=5$, the holes are in a gas-like state (similar to the electrons), at $M=50$ their structure resembles a liquid and, at $M=100$ and $800$, they are periodically arranged in space. Thus, between $M=100$ and $50$ the holes crystallize. The figure also clearly shows the mechanism of this quantum melting process: With reduction of $M$ the individual hole wave functions grow continuously until, at $M=M^{cr}$, they exceed 
a critical size which allows for tunneling between lattice sites giving rise to hole delocalization, i.e. crystal melting. Vice versa, increase of $M$ reduces the spatial extension of each hole which stabilizes the crystal (at $M=800$, they shrink to a dot, 
see Fig.~\ref{snap}). 

The crystallization transition is further supported by the behavior of the pair distribution functions, $g_{ab}$, shown in Fig. \ref{pdf}. At $M\le 50$ (upper two panels) the h-h functions (black curves) has only a single peak like in a liquid. However, for $M=100$, $g_{hh}$  exhibits periodic oscillations with a deep first minimum, typical for a crystal. The crystal exists only at low temperature, an  increase of $T$ by a factor of 2 (from the lower right figure to the lower left) causes thermal melting. Further, the e-e and e-h pair distributions allow us to understand the behavior of the electrons: In the hole crystal phase the electrons exhibit periodic density modulations indicating the formation of Bloch waves (band structure) with increased concentration at distances smaller than $0.5a_B$. Finally, we computed the relative distance fluctuations of the holes as a function of mass ratio at $T_e=0.64$ and $r_{se}=0.63$.  They show a strong increase around $M=80\dots 100$ typical for a solid-liquid transition (not displayed). 
\vspace{-0.35cm}

\begin{figure}[h]
\includegraphics[width=7cm,clip=true]{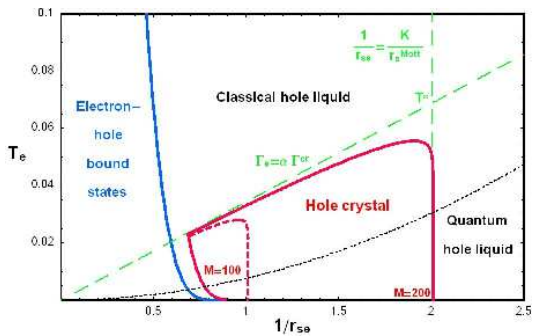}
\vspace{-0.35cm}
\caption{(Color online) Phase diagram of a two-component plasma in the plane of dimensionless electron temperature $T_e$ and density 
parameter $1/r_{se}$. The blue line indicates the boundary of the Coulomb bound state phase 
given by $r_s^{Mott}(T_e)$. Above (below) the dotted black line holes are classical 
(degenerate). The red full (dashed) line is the boundary of the hole crystal for $\Theta=Z=1$ and $M=200$ ($M=100$) with the asymptotics given by 
Eqs. (\ref{cc},\ref{nmax}) (green dashed lines).
}
\label{phase1}
\end{figure}

{\it Phase diagram.} We now construct the generic phase diagram of the TCP which applies to all the different Coulomb crystals, see Fig. \ref{phase1}. Consider first the case of a hole crystal in semiconductors which is embedded into a dense Fermi gas of electrons. The holes behave classically above the black dotted line and quantum-mechanically below (this line is given by $n_h\Lambda_h^3=1$, where  $\Lambda_h=h/\sqrt{2\pi m_h k_B T_h}$ is the hole deBroglie wave length). The e-h bound state phase is shown in the left part and contains excitons and biexcitons and, eventually at low temperature, a Bose condensate, superfluid or an excitonic insulator \cite{wachter91}. On leaving this phase across its boundary (given by the blue line $r_{se}(T_e)=r_s^{{\rm Mott}}(T_e)$) the fraction of bound states rapidly vanishes in favor of unbound e-h pairs with the holes showing liquid-like behavior.  Upon further compression (at temperatures $T_e<T^*$) the hole liquid crystallizes, provided $M\ge M^{cr}$. At the density $n^{(2)}$ quantum melting of the crystal is observed (vertical dashed green line). The entire hole Coulomb crystal phase for M=200 is marked by the full red line in Fig. \ref{phase1}.

Now, how is the hole crystal in semiconductors related to the Coulomb crystals in classical and astrophysical plasmas mentioned in the introduction? To answer this question we investigate the dependence of the stability of the crystal phase on the three asymmetry parameters. When $M$ is reduced the crystal phase shrinks (see the red dashed line corresponding to $M=100$)  until for 
$M=M^{cr}$ it vanishes. Vice versa, when $M$ becomes larger, both maximum density and temperature at which crystallization is possible, increase according to Eqs. (\ref{nmax}, \ref{tstar}). While in semiconductors in quasi-equilibrium $M$ is the only parameter which can be varied from one material to another, the diversity of ionic plasmas, on the other hand, offers additionally control of the charge and temperature ratios $Z$ and $\Theta$ in very broad ranges. By increasing $M$, $Z$ and $\Theta$ the crystal phase extends further towards higher density and temperature covering an ever increasing part of the temperature-density plane (with the exception of the bound state phase). Eventually this crystal phase will overlap with the known classical and astrophysical Coulomb crystals at low and high densities, respectively. Thus, indeed, the phase diagram in Fig. \ref{phase1} applies to all Coulomb crystals in two-component plasmas of electrons and point-like ions, independently of their physical origin. Of course, in different systems specific additional factors may exist. For example, phase separation or 
non-Coulombic bound states will modify the boundary of the bound state 
phase and the value of $n^{(1)}$, whereas band structure effects or extended ionic cores 
can modify the high-density behavior \cite{nmax_astro}. Finally, while our analytical results for the crystal phase are obtained neglecting e-h correlations (which is justified by the large mass ratio $M$), the simulations 
indicate that the electrons have a stabilizing influence, increasing the maximum temperature beyond $T^*$. These effects are beyond the present investigation and 
will be discussed elsewhere.

In summary, one of the most fundamental collective properties of matter -- Coulomb crystallization has been extended from the OCP model to the case of neutral systems with two oppositely charged components. Our analysis provides a common view and general quantitative bounds on the critical parameters for the existence of Coulomb crystals in a large variety of TCP, including dwarf stars, laser-cooled 
expanding plasmas, dusty plasmas and semiconductors. The critical parameters depend on the combined mass, charge and temperature asymmetry between the heavy and light component. Crystallization of protons and of holes in semiconductors is predicted.  Hole crystals should exist in materials with a mass ratio of about $80$ \cite{2d} and might be observable in rare earth semiconductors in
neutron scattering experiments. These hole crystals could serve as a valuable testing ground for quantum TCP crystals in general and for ion crystals in exotic stellar objects in particular.

\begin{acknowledgments}
This work is supported by the Deutsche Forschungsgemeinschaft via TRR 24, 
the RAS program No. 17 
%``Parallel calculations and multiprocessor computational systems''
and Award No. PZ-013-02 of the US CRDF.
\end{acknowledgments}


\begin{thebibliography}{99}
\bibitem{point} We distinguish plasma crystals from ionic crystals (such as NaCl) 
and metals the properties of which are essentially dictated by the electronic 
structure of the ionic cores.
\bibitem{nmax_astro} For ions with an internal structure the upper density limit may deviate from Eq. (\ref{nmax}) which refers to point charges as there compression may lead to structural transitions rather than melting.
\bibitem{2d} In two dimensions Eq.~(\ref{mcr}) yields $M^{cr}\approx 60$, where $r_{s2d}^{cr}\approx 74$ and $r_{s2d}^{{\rm Mott}}=1.2$, has been used ($r_{s2d}$ refers 
to the $2d$ exciton Bohr radius), but $r_{s2d}^{{\rm Mott}}$ still needs to be 
verified by simulations.
\end{thebibliography}
\end{document}